\documentstyle[prb,aps,amsfonts,amssymb,floats,epsfig]{revtex}
\begin{document}
\draft %
\twocolumn[\hsize\textwidth\columnwidth\hsize\csname@twocolumnfalse\endcsname 
\title{Low temperature ellipsometry of $\alpha^{\prime}$-NaV$_2$O$_5$}
\author{C.~Presura$^1$, D. van der ~Marel$^1$, A.~Damascelli$^2$,
 R.K.~Kremer$^3$}
\address{ $^1$Solid State Physics Laboratory, University of Groningen,
 Nijenborgh 4, 9747 AG Groningen, The Netherlands\\
 $^2$Laboratory for Advanced Materials, Stanford
  University, Stanford, CA 94305, USA\\
 $^3$Max-Planck-Institut f\"{u}r Festk\"{o}rperforschung,
  Heisenbergstrasse 1, 70569 Stuttgart, Germany}
\maketitle
\begin{abstract}
The dielectric function of $\alpha^{\prime}$-NaV$_2$O$_5$ was
measured with electric field along the {\em a} and {\em b} axes 
in the photon energy range 0.8-4.5 eV for temperatures down to 4K. 
We observe a pronounced decrease of the intensity of the 1 eV peak 
upon increasing temperature with an activation energy of about 25meV, 
indicating that a finite fraction of the rungs becomes occupied with 
two electrons while others are emptied as temperature increases. 
No appreciable shifts of peaks were found showing that the change 
in the valence state of individual V atoms at the phase transition is very small.
A remarkable inflection of this temperature dependence at the phase 
transition at 34 K  indicates that charge ordering is associated 
with the low temperature phase.
\end{abstract}
\pacs{PACS numbers: 78.40.-q, 71.35.-y, 75.50.-y}
\vskip2pc]
\narrowtext      
   
\begin{figure}
\centerline{\epsfig{figure=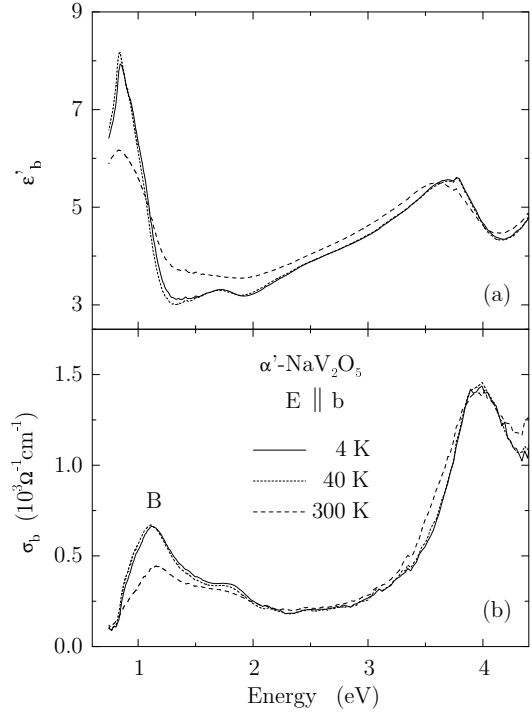,width=7cm, clip=}} \caption{
Real part  $\epsilon^{\prime}(\omega)$ of the complex dielectric
function (panel a), and optical conductivity $\sigma_1(\omega)$
(panel b), for $E\parallel b$.} 
\label{navob}
\end{figure}

\begin{figure}
\centerline{\epsfig{figure=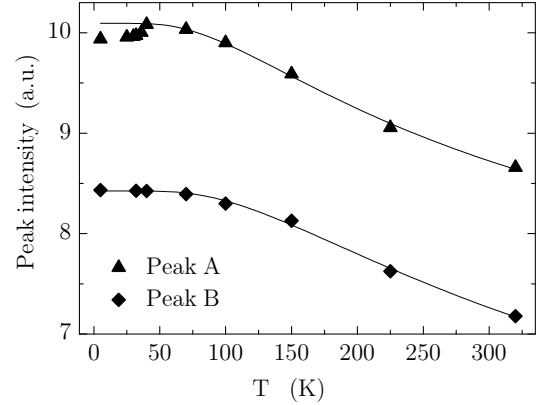,width=7cm,clip=}} \caption{
Intensity of the 1 eV peaks detected along the $a$
and $b$ axes (peak A and B, respectively), plotted vs. temperature.
Solid lines are fits to the formula $I(T)=I_0(1-f\,e^{-E_0/T})$.}
\label{navoint}
\end{figure}

\begin{figure}
\centerline{\epsfig{figure=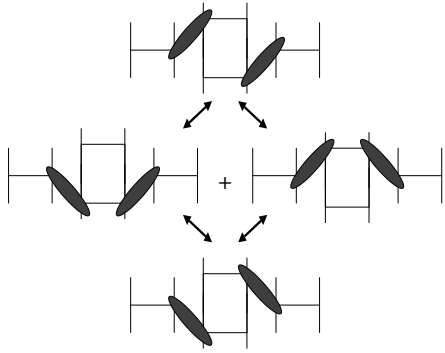,width=7cm,clip=}} \caption{
The correlated electron state in the low temperature phase is a superposition of the
four configurations displayed in the figure.}
\label{lowtemp}
\end{figure}

$\alpha^{\prime}$-NaV$_2$O$_5$ is subject of intensive research as
a result of its remarkable physical properties. The compounds
AV$_2$O$_5$ (A= Li, Na, Ca, Mg, etc.)\cite {Galy} all have the
same lattice structure, similar to that of 
V$_2$O$_5$. The structure can be described as two-legged ladders 
with VO$_5$ pyramids forming the corners arranged in 
two-dimensional sheets. In AV$_2$O$_5$ the A atoms
enter the space between the layers and act as electron donors for
the V$_2$O$_5$ layers. In $\alpha^{\prime}$-NaV$_2$O$_5$ 
the average valence of the V-ions corresponds to V$^{+4.5}$. X-ray 
diffraction indicates that at room temperature all V-ions are 
crystallographically equivalent\cite{meetsma,schnering,smolinski}.
At 35K a phase transition occurs, below which the following changes
take place: (i) A quadrupling of the unit cell\cite{isobe}, (ii) 
opening of a spin gap \cite{isobe}. In addition there are several
experimental hints for a charge redistribution below the phase
transition {\em e.g.} unaccounted for changes in
engropy\cite{buechner}, splitting of the V-NMR lines\cite{ohama},
inequivalent V-sites observed with XPS\cite{ludecke}. In this
paper we investigate the charge redistribution using optical
spectroscopic ellipsometry as a function of temperature. In
our spectra we observe clear indication of a strong charge
redistribution between the rungs of the ladders at elevated
temperature, which at the same time provides a channel for
electrical conductivity with an activation energy of about 25 meV.
We also report a remarkable inflection of the temperature dependence
at the phase transition, which we interpret as an inflection of the
charge redistribution process due to a particular 
correlated electronic state in which the charge and spin 
degrees of freedom are frozen out simultaneously.
 %
%
The crystal (sample CR8) with dimensions of 
approximately 2, 3 and 0.3 mm along the
{\em a}, {\em  b}, and  {\em  c} axes respectively, was mounted
in a UHV optical cryostat in order to prevent the formation of ice
on the surface. The pressure was about 10$^{-8}$ mbar at 300K and
reached 10$^{-9}$ mbar at 4K.
%
We performed ellipsometric measurements on the (001) surfaces 
of the crystals both with the plane of incidence of the light
along the {\em  a} and the {\em  b} axis. An
angle of incidence $\Theta$ of $80^{0}$ was used 
in all experiments. In Ref.\cite{nacavo} we 
describe the details of the procedure followed to obtain
$\epsilon(\omega)$. 
\\
The room temperature results are in general agreement 
with previous results \cite{damascelli,golubchik} using 
Kramers-Kroning analysis of reflectivity data. Along 
the a-direction we observe a peak at 0.9 eV with a shoulder
at 1.4 eV, a peak at 3.3 eV and the slope of a peak above 4.2 eV,
outside our spectral window. A similar blue-shifted sequence is 
observed along the b-direction. The 1eV peak drops rather sharply
and extrapolates to zero at 0.7 eV. However, weak absorption has 
been observed within the entire far and mid-infrared 
range\cite{smirnov,damascelli}. The strong optical absorption 
within the entire visible spectrum causes the characteristic 
black appearance of this material.  
Based on the doping dependence of the optical spectra of 
Na$_{1-x}$Ca$_x$VO$_5$ we established in Ref.\cite{nacavo} the
assignments made in Refs.\cite{smolinski,phorsch,damascelli}, namely that
the peak at around 1 eV along the a and b direction is due to transitions 
between linear combinations of V 3d$_{xy}$-states of the two V-ions forming the 
rungs\cite{smolinski,phorsch,damascelli}. In Refs.\cite{smolinski}
and \cite{phorsch} even and odd combinations were considered.
The 0.9 eV peak in $\sigma_{a}(\omega)$ (peak A) would then correspond 
to the transition from V-V bonding to antibonding combinations on the same 
rung\cite{damascelli}. In Ref.\cite{damascelli} 
this model was extended to allow lop-sided linear combinations of the 
same orbitals, so that the 0.9 eV peak then is a transition between 
left- and right-oriented linear combinations. The work presented
in Ref. \cite{nacavo} definitely rules out the assignment of these
peaks to crystal field-type V$d-d$ transitions proposed in
Refs. \cite{golubchik,long}.
\\     
The 1.1 eV peak in $\sigma_{b}(\omega)$ (peak B)
involves transitions between neighboring rungs
along the ladder. 
As a result of the correlation gap in the density of states,
the optically induced transfer of electrons
between neighboring rungs results in a final state with one
rung empty, and a neighboring rung doubly occupied, in
other words, an electron hole pair consisting of a hole in
the band below E$_F$, and an electron in the empty state above
E$_F$. Note that the final state
wavefunction is qualitatively different from the
on-rung bonding-antibonding excitations considered above (peak A),
even though the excitation energies are the same\cite{nacavo} : it involves one
rung with no electron, and a neighboring rung with one electron
occupying each V-atom. We associate the lower energy of
peak A compared to peak B with the attractive electron-hole Coulomb
interaction, favoring on-rung electron-hole pairs.
Optical transitions having values below 2eV
were also seen in V$_6$O$_{13}$ and VO$_2$. In  V$_2$O$_5$ they have
very small intensities, and were attributed to defects \cite{kenny}.
The peak at 3.3 eV in $\sigma_{a}(\omega)$ we could attribute to
a transition from the $2p$ orbital of oxygen to the antibonding
level within the same V$_2$O cluster\cite{nacavo}.
\\
%
%
Let us now address the temperature dependence of the spectra. 
Perhaps most striking of all is the fact that the {\em peak positions}
turn out to be temperature {\em in}dependent throughout the entire
temperature range. This behavior should be contrasted with the
remarkable splitting of the V-NMR lines in two components below
the phase transition\cite{ohama}. It has been suggested \cite{smolinski}
that the T$_c$ marks the transition                               
from a high temperature phase where every rung is occupied with
an electron residing in a H$_2^+$ type bonding orbital (formed by
the two V3d$_{xy}$ orbitals), to a low temperature phase, where the system
is in a charge ordered state ({\em e.g.} the zigzag ordered
state\cite{mostovoy,poilblanc} with
an alternation of V$^{4+}$ and V$^{5+}$ states, or, as suggested in
Refs. \cite{ludecke,palstra} with V$^{4+}$/V$^{5+}$ ladders and
V$^{4.5+}$/V$^{4.5+}$ ladders alternating).
In Refs.\cite{damascelli,cuoco}
estimates have been made of the potential energy difference between the 
left and righthand V-sites on the same rung, in order to reproduce
the correct intensity and photon-energy of the 1 eV peak along $a$,
as well as producing a V$^{4.9+}$/V$^{4.1+}$ distribution between
left and right. This turned out to be $\Delta=V_{L}-V_{R}\approx 0.8 eV$, with
an effective hopping parameter $t_{\perp}\approx 0.4 eV$.
\\
To have V$^{4.5+}$/V$^{4.5+}$ above and V$^{4+}$/V$^{5+}$ below 
the phase transition, requires than that the potential energy
difference changes from $\Delta=0.8 eV$ below T$_c$ to $0$ at and above
$T_c$. As a result the "1 eV peak" would shift from 0.89 eV to 0.8 eV
in the temperature interval between 0 and 34 K, and would remain constant
above $T_c$. The observed shift is less than 0.03 eV within the entire
temperature interval, and less than 0.01 eV between 0 and 34 K. 
This suggests that the change in $\Delta$ (and consequently 
the charge of the V atoms) at the phase transition is very small. In fact,
a change of $\Delta$  from $0.1 eV$ to $0$ across T$_c$, 
compatible with the experimental
results, would yield a change in the valence state from
V$^{4.44+}$/V$^{4.56+}$  to V$^{4.5+}$/V$^{4.5+}$ between
0 and 34 K, which is an almost negligible effect.
Thus we conclude that, irrespective of the possible charge configurations
V$^{4.5+}$/V$^{4.5+}$ or V$^{5+}$/V$^{4+}$, the changes in the charges
of the V atoms at the phase transitions are very small (smaller than 0.06{\em e}).
\\
\begin{figure}
\centerline{\epsfig{figure=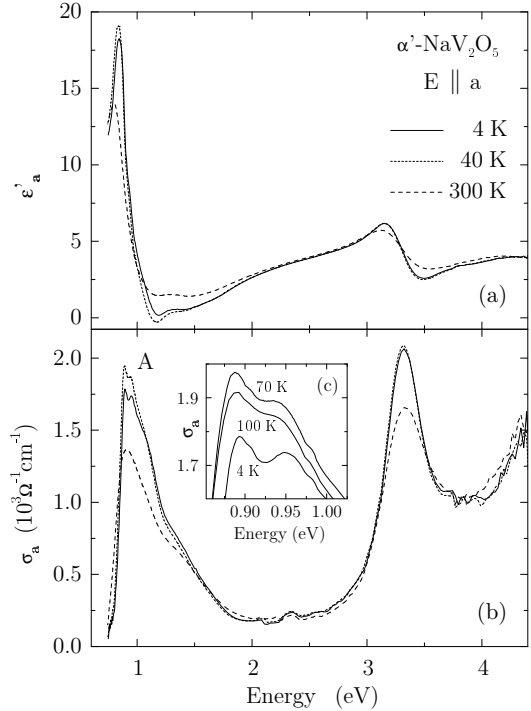,width=7cm,clip=}} \caption{
Real part  $\epsilon^{\prime}(\omega)$ of the complex dielectric
function (panel a), and optical conductivity $\sigma_1(\omega)$
(panel b), for $E\parallel a$.}
\label{navoa}
\end{figure}
As we can see in Fig. \ref{navoa}b and \ref{navob}b, there
is a strong decrease of the {\em intensity} of the peaks A and B with
the increase of the temperature. The spectral weight for both
cases is not transferred to low frequencies\cite {damascelli}. The
spectral weight of the B peak seems to be recovered up to and
above 4eV. The spectral weight of the 3.3eV peak in the  {\em  a}
direction is recovered also in the nearby high frequencies
\cite{konstantinovic,long}, whereas the intensity of the A peak
 seems to be recovered at even higher photon energy, probably 4.5 eV
\cite{konstantinovic}. The evolution of the 1 eV peaks can be seen
from Fig. \ref{navoint}, where the integrated intensities in
$\sigma_1(\omega)$ from 0.75 eV to 2.25 eV were plotted as a
function of temperature. The data fitted with the formula
$I(T)=I_0(1-f\,e^{-E_0/T})$ gave $f$=0.35 and $E_0$=286K for the
{\em  a} direction and $f$=0.47 and $E_0$=370K for the  {\em  b}
direction. From the fits we see that the activation energy $E_0$
is about 25meV, which is very small for the frequency range of the
peaks. A decrease of the intensity of the A peak takes place below
the phase transition, but otherwise there are no features related
to it.  The splitting of the A
peak of about 55meV (Fig.\ref{navoa}c) exists even at 100K. Judging from its
sharp shape and the value of splitting, it can be attributed to a
phonon side-band. 
\\
Band structure calculations have indicated that the  d$_x$$_y$ orbitals are
well separated from the other d orbitals \cite{smolinski} and ESR
experiments have led to g-values which indicate the complete
quenching of the orbital momentum \cite{lohmann}. There are then
no other low-lying crystal levels, about 25 meV above the
ground state, to play a role in the temperature dependence
behavior of the A peak.
Comparing the doping dependence of the A peak in Ca$_x$Na$_{1-x}$V$_2$O$_5$
\cite{nacavo} and  the high temperature dependence from Fig.\ref{navoa}b
we see that the two behaviors 
resembles, presenting no shifting or splitting. But, as discussed in Ref \cite{nacavo},   
the intensity of the A peak decreases upon doping because doping induces 
doubly occupied rungs. The same mechanism can then be responsible for the
decrease of the intensity of the A peak with increasing temperature.
The bonding-antibonding transition (A peak) on
the rung will have a reduced intensity, as there are fewer singly
occupied rungs, as in the case of Ca$_x$Na$_{1-x}$V$_2$O$_5$\cite{nacavo}. 
The transitions on the doubly occupied rungs are
at an energy $U$, around 4 eV, with a factor $4t_{\perp}/U$
reduction of the original spectral weight\cite{nacavo}. The activation energy
of 25meV would be then the energy required to redistribute the
electrons between the rungs, either on the same ladder, or between
different ladders. Eventually, at very high temperatures, only
half of the rungs would be occupied with one electron, so the
intensity of the A peak would be at half the low temperature value
($f=0.5$ in the fitting formula of Fig.\ref{navoint}). 
\\
At first glance the processes leading to partial emptying of
rungs, while doubly occupying others, seem to be of the order of
the energy of peak B (1eV), which corresponds exactly to such a
process and one may wonder how a low energy scale could exist.
However, processes involving the {\em collective} motion of charge
can be at a much lower energy than the single particle charge
transfer, as a result of short range (nearest neighbor) Coulomb
interactions. An example of such a collective mode is the zig-zag 
ordering \cite{mostovoy,cuoco} involving an (almost) soft
charge mode for $k$ at the Brillouin zone boundary. 
These {\em charge} modes, because $k$ is at the BZ boundary, can appear
only indirectly ({\em e.g.} phonon assisted) in $\sigma(\omega)$, and
therefore are at best weakly infrared active. Under favorable
conditions the {\em spin} degrees of freedom\cite{damascelli} in
addition result in a weak but finite $\sigma(\omega)$. 
Another way in which the electrons can move from one rung to another is by
forming topological defects, such as domain cells separating
charge ordering domains. Macroscopically this could lead to double
occupancy of some rungs and emptying others.
\\
In fact, even though the optical gap is 1eV,
there are experiments which indicate charge
degrees of freedom at a much lower energy. Resistivity measurements
yielded an energy gap ranging from 30meV at lower temperatures to
75 meV at high temperatures \cite{hemberger}.
The dielectric loss $\epsilon''$ for frequency
of 16.5 GHz along b direction is rather
constant up to 150 K and then increases very
rapidly above 200K \cite{poirier}
(so that the microwave signal is lost at room temperature), meaning that
an absorption peak could start to evolve at 200K for very low frequencies.
A low frequency continuum was observed near 25 meV with infrared
spectroscopy \cite{damascelli} and at 75 meV with Raman
spectroscopy \cite{golubchik}.
Also infrared measurements \cite{damascelli}
found that $\sigma_{1,a}$ increases with increasing temperature.
%
%
\\  
The suppression of intensity below the phase transition in this
context (Fig.\ref{navoint})
seems to mark a redistribution of charge which
is associated with the spin gap. X-ray diffraction indicates
that the superstructure below T$_c$  consists of a group of 
4 rungs \cite{palstra}: 2 neighboring rungs of the central ladder,
1 on the left-hand and 1 on the right-hand ladder. The presence
of a spin-gap indicates that the 4 spins of this structural 
unit form an $S=0$ state below T$_c$. To account for 
the absence of a change of the valence of the V atoms at the 
phase transition, as well as for the this slight doubly occupancy below the
phase transition, the following scenario can be put forward. Below T$_c$
the structure would be formed by singlets (see Fig. \ref{lowtemp}). 
A possible arrangement, which is motivated by the observed crystal structure
at low temperature\cite{ludecke} is indicated in Fig. \ref{lowtemp}. It 
corresponds to mainly two degenerate configurations involving one electron
on each rung (top and bottom), engaged in a singlet formed of two electrons on
nearest neighbor V-positions on {\em different} ladders. 
These diagonal singlets were originally proposed by Chakraverty {\em et al.}\cite{chakra} 
for Na$_{0.33}$V$_2$O$_5$. The middle two configurations are at higher energy
states of order 1 eV, hence they are only slightly mixed in. Because the latter
configurations have one empty and one doubly occupied rung, the intensity
of the A and B peak should again be reduced, as was discussed for temperatures
above the phase transition. The reduced
intensity in our spectra  below $T_c$ then reflects the amount of singlet
character involving doubly occupied rungs. 
Passing the phase transition the coherence of this state would vanish. 
This would result in a random configuration with an average
valence of +4.5 for the V atoms, and also a spin susceptibility for
high temperature phase due to appearance of some free spins.
The nature of the weak charge-redistribution which we observe
at low temperature would then be manifestly {\em quantum mechanical}.
\\   
In conclusion, we have measured the temperature dependence behavior
of the dielectric function along the
{\em  a} and {\em  b} axes of $\alpha^{\prime}$-NaV$_2$O$_5$
in the photon energy range 0.8-4.5 eV for temperatures down to 4K.
No appreciable shifts of the 1 eV peaks were found,
thus showing that the change in the valence state of the V atoms at 
the phase transition is very small (smaller than 0.06{\em e}).
A strong decrease of the 1 eV peaks with increasing temperature was observed. 
We assigned this temperature dependence behavior to collective
charge redistribution, namely the redistribution of the electrons
among the rungs resulting in double occupation of some rungs as
temperature increases, with an activation energy of about 25meV.
Below the phase transition, a small but sharp decrease of
intensity of the  0.9 eV peak in $\sigma_{a}(\omega)$ was found.
It was attributed to a finite probability of having, in the singlet state
below T$_c$, configurations with electron pairs occupying the same rung.
\\ 
We gratefully acknowledge T.T.M. Palstra, J.G. Snijders, M. Cuoco,
and A. Revcolevschi for stimulating discussions. 
This investigation was supported by the Netherlands Foundation for
Fundamental Research on Matter (FOM) with financial aid from the
Nederlandse Organisatie voor Wetenschappelijk Onderzoek (NWO). 
%

%
%
%
%
%
%
%
\end{document}